\documentclass[twocolumn,prl,aps,superbib,tightenlines,floatfix]{revtex4}
\usepackage{amsfonts}
\usepackage{amsmath}
\usepackage{amssymb}
\usepackage{graphicx}

\begin{document}

\title{Realization of arbitrary single-qubit gates through control of
spin-orbit couplings in semiconductor nanowires}
\author{S. J. Gong and Z. Q. Yang\cite{ZY}}
\affiliation{Surface Physics Laboratory (National Key Laboratory), Fudan University,
Shanghai, 200433, China.}

\begin{abstract}
We propose a theoretical scheme to realize arbitrary single-qubit
gates through two simple device units: one-dimensional semiconductor
wires with Dresselhaus spin-orbit coupling (SOC) and Rashba SOC,
separately. Qubit information coded by the electron spin can be
accurately manipulated by the SOC when crossing the semiconductor
wire. The different manipulative behaviors in Dresselhaus and Rashba
wires enable us to make the diverse quantum logic gates.
Furthermore, by connecting the Dresselhaus and Rashba units in
series, we obtain a universal set of single qubit gates: Hadamard,
phase, and $\pi /8$ gates, inferring that an arbitrary single qubit
gate can be achieved. Because the total transmission is satisfied in
the two device units, all the logic gates we have obtained are
lossless. In addition, a ballistic spintronic switch is proposed in
the present investigation.

\medskip PACS Numbers: {71.70.Ej, 85.35.Be, 03.67.-a}
\end{abstract}

\keywords{Qubit gates, spin-orbit coupling, Rashba, Dresselhaus}
\maketitle

Quantum logic gate is the basic building block of quantum computer, which
has attracted considerable attention recently. A variety of physical
implementations have been proposed to realize the quantum gates \cite%
{Nielsen,Yamamoto,Step}, among them the solid-state (especially
nanoelectronic) implementations, are regarded to be one of the most suitable
candidates to realize truly large-scale quantum computer in reality.
Meanwhile, the qubit information coded by the spin degrees of freedom in
solid-state system, is believed to be much more robust and stable than
charge qubit, for having relatively long coherence time \cite{Kane,Band}.
Based on these two considerations, a large class of spin-based logic gate
schemes have been proposed in semicondutor nanostructures, mainly in
semiconductor quantum dots \cite{Step,Band,Loss}. The one and only exception
is the scheme presented by F\"{o}ldi \textit{et al} \cite{Foldi}, who first
concretely designed the spintronic single-qubit gates using one-dimensional
semiconductor rings, based on the fact that the precession of the eletron
spin can be accurately manipulated by the Rashba spin-orbit coupling (SOC)
\cite{Rashba} without the requirement of external magnetic field \cite%
{Datta,Mire}. It is thus feasible in principle to carry out all-electrical
quantum computation in future \cite{Popescu}. However, the functions of the
logic gates they proposed were tuned a little bit complicated by changing
both the external electric field and the geometries of the units. That is,
in addition to the relatively easy tuning of Rashba SOC by electric field,
several rings with different sizes and different geometry angle\textbf{s}
must be fabricated, which greatly reduces the flexibility and unitary of the
device assemble.

Except the Rashba interaction, there is another typical SOC, i.e.
Dresselhauls SOC \cite{Dresselhaus} exiting in the semiconductor sample,
which appears as a result of bulk\ \cite{Dresselhaus} and interface \cite%
{Dyak} inversion asymmetries. Its strength had been considered not to be
tuned as conveniently as that of Rashba interaction. Recent theoretical \cite%
{Wu} and experimental \cite{Kato2} work showed the Dresselhaus spin
splitting could be effectively adjusted via strain existing in the
structure. Schliemann and Loss \cite{Schliemann} predicted that the
Dresselhaus coefficient in a quantum well might also be conveniently altered
by an electric field. Therefore, it is possible to carry out practical ways
in experiments to control the strength of Dresselhaus interaction.

In the present work, we propose a theoretical scheme to implement a
universal set of single-qubit gates in semiconductor nanowires, based on
both Rashba and Dresselhaus SOCs. The basic device units are two
quasi-one-dimensional electron gas (1DEG) systems with either Dresselhaus or
Rashba SOC, sandwiched between two leads. We find both the Dresselhaus and
Rashba units can serve as lossless quantum gates, but the properties of the
gates are different resulting from the different SOC mechanisms. By
connecting the two device units in series, a universal set of single-qubit
gates: Hadamard, phase, $\pi /8$ gates \cite{Nielsen}, can be implemented,
inferring an arbitrary single qubit gate can be obtained in our scheme. The
important arbitrary phase gate \cite{Nielsen} can be obtained by tuning the
strength of Dresselhaus interaction, which can greatly optimize the circuits
when building quantum computer. In addition, a ballistic spintronic switch
is suggested.

\begin{figure}[htbp]
\includegraphics*[width=8.0cm]{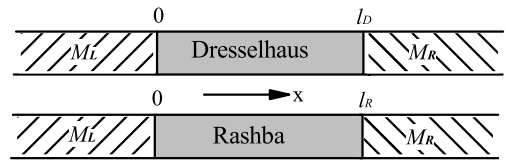}
\caption{The geometries of two device units of single-qubit gates made from
Dresselhaus (above) or Rashba (below) SOCs. $M_{L\text{ }}$and $M_{R}$
represent the left and right electrodes, respectively.}
\end{figure}

Figure 1 shows the schematics of the two device units: Dresselhaus-type
(above) and Rashba-type nanowires (below). When the cross-section of the
nanowire is sufficiently small, only electronic motion in the longitudinal
direction is of interest, i.e., the system becomes quasi-one-dimensional
\cite{note1}. The SOC region along $x$\ direction is sandwiched between two
leads. An electron wave is injected from the left lead to the right one
crossing the middle SOC region. The Schr\"{o}dinger equations in the three
regions are solved separately. The Hamiltonian\textbf{s} in the Dresselhaus-
and Rashba-type nanowires can be respectively described by\textbf{\ }%
\begin{eqnarray}
H_{D} &=&-\frac{\hbar ^{2}}{2m^{\ast }}\nabla _{x}^{2}-\dfrac{\alpha _{D}}{%
\hbar }\sigma _{x}p_{x}+U[\delta (x)+\delta (x-l_{D})], \\
H_{R} &=&-\frac{\hbar ^{2}}{2m^{\ast }}\nabla _{x}^{2}-\dfrac{\alpha _{R}}{%
\hbar }\sigma _{y}p_{x}+U[\delta (x)+\delta (x-l_{R})],
\end{eqnarray}%
where $m^{\ast }$\ is the effective mass of electrons and $\sigma _{x},$ $%
\sigma _{y}$ are the Pauli matrices. The coefficients $\alpha _{D}$\ and $%
\alpha _{R}$ express the Dresselhaus and Rashba strength\textbf{s},
respectively.\ We choose $\alpha _{D/R}=0$\ in leads, and $\alpha _{D/R}\neq
0$\ in the middle SOC region. To model the elastic scattering that usually
occurs at the two interfaces between the leads and the SOC region, we follow
the previous work to include a $\delta $-function\ potential with the height
of $U$, at positions of $x=0$\ and $l_{D/R\text{ }}$\cite{Blonder, Matsuyama}%
. The parameters $l_{D\text{ }}$and $l_{R\text{ }}$indicate the\textbf{\ }%
lengths\textbf{\ }of the Dresselhasu and Rashba regions, respectively. A
dimensionless parameter $z=(U/\hbar )\sqrt{2m^{\ast }/E}$ is introduced to
represent the strength of the interfacial scattering, where $E$ is the
incident electron energy. No spin-flip across the interfaces is assumed in
the calculation \cite{Blonder, Matsuyama}.

The wave functions in the left (source) and right (drain) leads can be
written as $\Psi _{l}(x)=\left(
\begin{array}{c}
\alpha \\
\beta%
\end{array}%
\right) e^{ik_{x}^{M}x}+\left(
\begin{array}{c}
R_{1} \\
R_{2}%
\end{array}%
\right) e^{-ik_{x}^{M}x}$ and $\Psi _{r}(x)=\left(
\begin{array}{c}
T_{1} \\
T_{2}%
\end{array}%
\right) e^{ik_{x}^{M}x},$ respectively, where $k_{x}^{M}$ is the wave vector
in the leads. The spinors $\left(
\begin{array}{c}
\alpha \\
\beta%
\end{array}%
\right) $, $\left(
\begin{array}{c}
R_{1} \\
R_{2}%
\end{array}%
\right) $, and $\left(
\begin{array}{c}
T_{1} \\
T_{2}%
\end{array}%
\right) $ respectively represent the spin states of the incident,
reflective, and transmission waves. The coefficients $\alpha ,\beta $ are
arbitrary complex numbers and satisfy $|\alpha |^{2}+|\beta |^{2}=1$. If the
two coefficients $R_{1}$ and $R_{2}$ in the reflective wave are known, the
reflective probability $R$ can be obtained through $%
R=|R_{1}|^{2}+|R_{2}|^{2}.$ The spin state becomes $\left(
\begin{array}{c}
T_{1} \\
T_{2}%
\end{array}%
\right) $ after being tuned by the middle SOC interaction. The transmission
coefficient $T$ of the outgoing wave can be obtained from $%
T=|T_{1}|^{2}+|T_{2}|^{2}.$ According to the conservation of the particle
number, $T+R=1$ is guaranteed in all the calculations.

The degenerate up- and down-spin states at $\pm k_{x}^{M}$ in electrodes
(corresponding to the incident energy $E$) will lift into four different
wave vectors $\pm k_{x\text{ }}$and $\pm k_{x}^{\prime }$ in the middle
region due to the SOC interaction. Therefore, the up- and down-spin wave
functions in the middle SOC region can be expanded with the four
corresponding eigenspinors. The wave function in this region can be
expressed as $\Psi _{mid}=\left(
\begin{array}{l}
\sum\limits_{j=1-4}{C(}\mathit{k}_{x}^{j})e^{ik_{x}^{j}x} \\
\sum\limits_{j=1-4}{D(\mathit{k}_{x}^{j})}e^{ik_{x}^{j}x}%
\end{array}%
\right) .$ At a fixed $\mathit{k}_{x}^{j}$, the coefficient ${C(}\mathit{k}%
_{x}^{j})$ is related to ${D(}\mathit{k}_{x}^{j})$ by the eigenspinors for
infinite SOC system. It is found that for the two kinds of SOCs, the
eigenspinors for the infinite systems are different. There are four
independent coefficients to be determined in the SOC region. They together
with another four coefficients in the two leads ($R_{1,}$ $R_{2,\text{ }%
}T_{1},$ $T_{2}$) can be solved by boundary conditions at the two interfaces
normal to $x$ direction \cite{Yao}. After the equations are solved, we can
get the spin states of outgoing wave and establish the relations between the
incoming and the outgoing spin states by introducing a transformation matrix
$G(2)$: $\ \ $%
\begin{equation}
\ \left(
\begin{array}{c}
T_{1} \\
T_{2}%
\end{array}%
\right) =G(2)\left(
\begin{array}{c}
\alpha  \\
\beta
\end{array}%
\right) .
\end{equation}%
In Dresselhaus device, $G(2)=T\cdot G_{D}(2)$, and in Rashba one, $%
G(2)=T\cdot G_{R}(2)$, where $G_{D}(2)$ and $G_{R}(2)$ are unitary and
unimodular matrices, describing the spin transformation properties in
Dresselhaus and Rashba units, respectively. A possible global phase in the
form of $e^{i\theta \text{ }}$can be neglected in $G_{D}(2)$ and $G_{R}(2)$,
since the effect of the global phase is not observable in quantum
computation \cite{Nielsen}. To attain lossless qubit gates, the complete
transmission ($T=1$) must be satisfied.

In the calculation, the effective mass $m^{\ast }$ is set as $0.04$ $m_{e}$.
The lengths of the SOC regions $l_{D}$ and $l_{R}$, fixed at $200$ nm, are
restricted within ballistic region, i.e., $l_{D}$ and $l_{R}$ are smaller
than the phase-coherent length $l_{\phi }$, which is in the range of 0.4-1.0
$\mu $m \cite{Nitta,Nomura,Souma}). Ballistic samples with small dimensions%
\textbf{\ }are always desirable for future electronic device applications.
Fig. 2 shows the transmission coefficient for the Dresselhaus and Rashba
units, in which a moderate $\delta $-potential is considered. Here we note
that if the corresponding parameters of these two geometries are of the same
values, the totally same results of the transmission are obtained, which can
be ascribed to the same phase interference properties between the four waves
in the two kinds of SOC regions. Two bright camber areas corresponding to $%
T\simeq 1$ are clearly seen, within which the geometry can serve as a
lossless single-qubit gate. The energy width corresponding to the total
transmission is about $0.5$ meV (see the upper camber). It decreases a
little bit in the lower camber. This behavior shows that the bias applied on
the wires should be in linear region and within the magnitude of $0.5$ meV
to build the lossless qubit gates. Between the two total transmission
regions, there is a wide energy gap, in which the transmission coefficient
is too small to devise the logic gates.
\begin{figure}[htbp]
\includegraphics*[width=8cm]{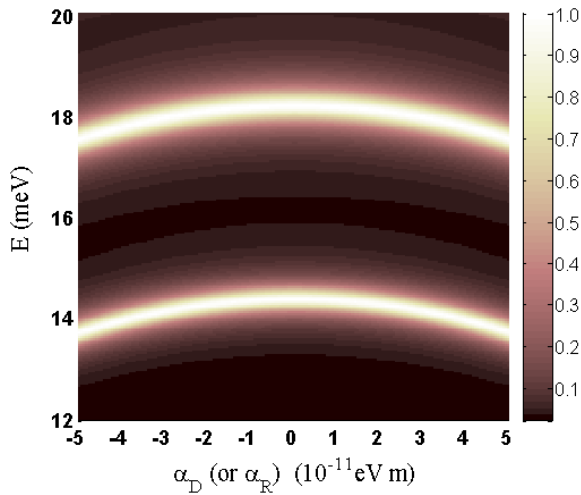}
\caption{Contour plots of the transmission coefficient $T$ as a function of
the incident energy ($E$) and the SOC strength ($\protect\alpha _{D}$ or $%
\protect\alpha _{R}$).}
\end{figure}

Except the SOC strength and the incident energy, the height of the $\delta $
potential at the two interfaces also significantly influences the
transmission of the electron. Figure 3 provides the transmission coefficient
as a function of the SOC strength ($\alpha _{D}$ or $\alpha _{R}$) under
various magnitudes of $\delta $ potential, where the incident energy is
fixed at $18$ meV. When $\alpha _{D/R}=0$, the total transmission can only
occur at certain $\delta $ potential height $z_{0}$ $=0.017$. If $z$
deviates from $z_{0}$ (either larger or smaller), the transmission
decreases, indicating that the resonant-like transmission occurs between the
two $\delta $ potentials with the height of $z_{0}$. When the potential
barrier is fixed at $z_{0}$, the total transmission is obtained in a wide
range of Rashba strength, which means a great number of logic gates can be
realized there. Increasing the height of the potential, two narrower peaks
corresponding to $T\simeq 1$ appear. In practice case, we certainly hope to
find the situation corresponding to the solid curve.

\begin{figure}[htbp]
\includegraphics*[width=8.0cm]{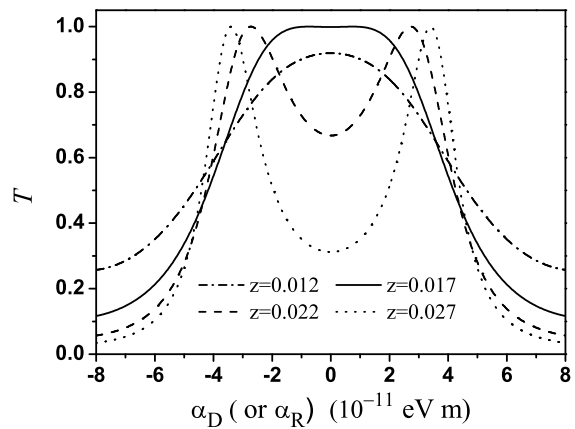}
\caption{The transmission coefficient as a function of the SOC strength ($%
\protect\alpha _{D}$ or $\protect\alpha _{R}$). The incident energy $E=18$
meV.}
\end{figure}

The spin transformation properties in Dresselhaus and Rashba device units
are found to be different: $G_{D}(2)$ $=$ $\left(
\begin{array}{cc}
\cos \frac{\theta _{D}}{2} & i\sin \frac{\theta _{D}}{2} \\
i\sin \frac{\theta _{D}}{2} & \cos \frac{\theta _{D}}{2}%
\end{array}%
\right) $, $G_{R}(2)$ $=$ $\left(
\begin{array}{cc}
\cos \frac{\theta _{R}}{2} & \sin \frac{\theta _{R}}{2} \\
-\sin \frac{\theta _{R}}{2} & \cos \frac{\theta _{R}}{2}%
\end{array}%
\right) $,\ where the parameters $\theta _{D}=\frac{2m^{\ast }\alpha
_{D}l_{D}}{\hbar ^{2}}$ and $\theta _{R}=\frac{2m^{\ast }\alpha _{\theta
}l_{\theta }}{\hbar ^{2}}$. They express the rotation angles of spinors of
electrons after crossing the spin-orbital coupling regions. It is reasonable
that the rotation angle is proportional to the strength of the SOC and the
length of the SOC region\cite{Mire}.

Various qubit gates can be realized by the above two device units. Clearly,
the non-trivial NOT gate \cite{Nielsen} can be implemented through
Dresselhaus device unit, if $\theta _{D}=\pi $ is satisfied at the base of $%
T=1$. For example, if the Dresselhaus device parameters are set as: $%
l_{D}=200$ nm, $E=18$ meV, $z=0.017$, the NOT gate can be approximately
obtained when $\alpha _{D}\simeq 1.496\times 10^{-11\text{ }}$eVm. Besides
the NOT gate, another two unitary matrices $\frac{1}{\sqrt{2}}\left(
\begin{array}{cc}
1 & i \\
i & 1%
\end{array}%
\right) $ and $\left(
\begin{array}{cc}
\cos (\frac{\pi }{8}) & i\sin (\frac{\pi }{8}) \\
i\sin (\frac{\pi }{8}) & \cos (\frac{\pi }{8})%
\end{array}%
\right) $ can be realized in Dresselhaus device by tuning the value of $%
\alpha _{D}$, and will play a critical role in devising the
important phase and $\pi /8$ gates, which will be discussed in
details later .

For Rashba device unit, if the spin rotation angle $\theta _{R}=\frac{\pi }{2%
}$ and $T=1$, we can obtain the transformation matrix of $\frac{1}{\sqrt{2}}%
\left(
\begin{array}{cc}
1 & 1 \\
-1 & 1%
\end{array}%
\right) $. This is a pseudo-Hadamard gate because it lacks the correct
determinant $(-1)$ of genuine Hadamard.\textbf{\ }The similar conclusion has
also been drawn in quantum Rashba-type ring \cite{Foldi}. If the Rashba
device are modeled with the parameters: $l_{R}=200$ nm, $E=18$ meV, and $%
z=0.017$, the pseudo-Hadamard can be attained when $\alpha _{R}$ is
fixed at $7.482\times 10^{-12\text{ }}$ eV m. Tuning $\alpha _{R}$
to let $\theta _{R}=\pi $, we get the pseudo-NOT gate $\left(
\begin{array}{cc}
0 & 1 \\
-1 & 0%
\end{array}%
\right) $. In addition, we find the identity matrix can be realized
with the condition of $\theta _{R}=2\pi $. This means that the
Rashba effect can be totally 'cancelled' and the electrons can reach
the right lead with its original spin state unchanged.

The product of the pseudo-NOT gate and the NOT gate gives the \textit{Z}
gate $\left(
\begin{array}{cc}
1 & 0 \\
0 & -1%
\end{array}%
\right) $. This means that the \textit{Z} qubit gate can be implemented by
connecting in series one Rashba device unit with one Dresselhaus unit.
Similarly, the genuine Hadamard gate $\frac{1}{\sqrt{2}}\left(
\begin{array}{cc}
1 & 1 \\
1 & -1%
\end{array}%
\right) $ \cite{Nielsen} can be achieved by connecting the pseudo-Hadamard
gate (made from the Rashba unit) with the NOT gate (made from the
Dresselhaus unit). We find the phase gate $\left(
\begin{array}{cc}
1 & 0 \\
0 & i%
\end{array}%
\right) $ and the $\pi /8$ gate $\left(
\begin{array}{cc}
1 & 0 \\
0 & e^{i\frac{\pi }{4}}%
\end{array}%
\right) $ can be decomposed as: $H\cdot $ $\frac{1}{\sqrt{2}}\left(
\begin{array}{cc}
1 & i \\
i & 1%
\end{array}%
\right) \cdot H$ and $H\cdot $ $\left(
\begin{array}{cc}
\cos (\frac{\pi }{8}) & i\sin (\frac{\pi }{8}) \\
i\sin (\frac{\pi }{8}) & \cos (\frac{\pi }{8})%
\end{array}%
\right) \cdot H$ ($H$ stands for the Hadamard gate), respectively. This
means that both of them can be realized by connecting three Dresselhaus and
two Rashba device units in series with certain sequence.\ The Hadamard,
phase, and $\pi /8$ gates constitute a universal set of single-logic gates,%
\textbf{\ }based on which any single quantum gates can be buil\textbf{t} to
arbitrary precision \cite{Nielsen}. A continuous phase gate $\left(
\begin{array}{cc}
e^{i\frac{\theta _{D}}{2}} & 0 \\
0 & e^{-i\frac{\theta _{D}}{2}}%
\end{array}%
\right) $ \cite{Nielsen}, decomposed as $H\cdot $ $G_{D}(2)\cdot H$, is
actually implemented by just varying the Dresselhaus strength. This is a
superior character compared with the previous work \cite{Foldi}, in which
the phase gate is tuned by changing the geometry angle.

In addition, from Fig. 3, we find that a ballistic spintronic switch can be
made by the Rashba device unit. The near complete transmission at certain
values of $\alpha _{R}$ corresponds to the \textquotedblleft
ON\textquotedblright\ of the switch. The near-zero transmission is regared
as \textquotedblleft OFF\textquotedblright . Therefore, by tuning the gate
voltage \cite{Nitta}, we can easily control the device to be
\textquotedblleft ON\textquotedblright\ or \textquotedblleft
OFF\textquotedblright . Since the sample size is in the ballistic region
with the length of $200$ nm, it may be used in future nano-sized electronic
devices.

In conclusion, a theoretical scheme is presented for realization of
arbitrary single-qubit gates based on two quasi-one-dimensional
semiconductor wires with Dresselhaus and Rashba interactions, separetely.
All the single-qubit gates would be realized possibly by using a series of
semiconductor nanowires with the same geometry and lead connection, which
simplifies the device manufacturing and is in favor of large scale
integration. In addition, a ballistic spintronic switch is proposed, whose
\textquotedblleft ON" and \textquotedblleft OFF" can be tuned conveniently
by varying external electric field.

The authors are grateful to Prof. X. Wang in Fudan Univ. for very helpful
discussion. This work was supported by\textbf{\ }the National Natural
Science Foundation of China with grant No.10304002, the Grand Foundation of
Shanghai Science and Technology (05DJ14003), PCSIRT, and the Fudan High-end
Computing Center.


\begin{thebibliography}{*}
\bibitem{note1} For example, the model can describe well a quasi-1DEG system
with width less than 20 nm, since the first confinement energy of the
geometry is found to be higer than 20 meV and the Fermi energy of a
semiconductor 2DEG is generally about or below 20 meV.
\end{thebibliography}
\end{document}